\documentclass[prx, twocolumn, showkeys]
{revtex4-2}
\usepackage{times}
\usepackage{bm}
\usepackage{graphicx}
\usepackage{amsbsy}
\usepackage{amsmath}
\usepackage{physics}
\usepackage{amsfonts}
\usepackage{amsthm}
\usepackage{natbib}
\usepackage[ruled,vlined]{algorithm2e}
\usepackage{enumitem}
\usepackage{color}
\usepackage{dcolumn}
\usepackage[utf8]{inputenc}
\usepackage{bm}  

\usepackage{lipsum}
\usepackage{subfigure}
\usepackage{caption}

\begin{document}

\theoremstyle{plain}
\newtheorem{theorem}{Theorem}
\newtheorem{lemma}[theorem]{Lemma}
\newtheorem{corollary}[theorem]{Corollary}
\newtheorem{proposition}[theorem]{$\eta_{r}$oposition}
\newtheorem{conjecture}[theorem]{Conjecture}

\theoremstyle{definition}
\newtheorem{definition}[theorem]{\gammaefinition}

\theoremstyle{remark}
\newtheorem*{remark}{Remark}
\newtheorem{example}{Example}

\title{The role of entanglement for enhancing the efficiency of quantum kernels towards classification}

\author{Diksha Sharma$^{1}$}
\author{Parvinder Singh$^{2}$}
\email{parvinder.singh@cup.edu.in}
\author{Atul Kumar$^{1}$}
\email{atulk@iitj.ac.in} 
\affiliation{$^{1}$Indian Institute of Technology Jodhpur, Rajasthan-342037, India}
\affiliation{$^{2}$Central University of Punjab Bathinda, Punjab-151001, India} 
\date{\today}

\begin{abstract}
Quantum kernels are considered as potential resources to illustrate benefits of quantum computing in machine learning. Considering the impact of hyperparameters on the performance of a classical machine learning model, it is imperative to identify promising hyperparameters using quantum kernel methods in order to achieve quantum advantages.  In this work, we analyse and classify sentiments of textual data using a new quantum kernel based on linear and full entangled circuits as hyperparameters for controlling the correlation among words. We also find that the use of linear and full entanglement further controls the expressivity of the Quantum Support Vector Machine (QSVM).  In addition, we also compare the efficiency of the proposed circuit with other quantum circuits and classical machine learning algorithms. Our results show that the proposed fully entangled circuit outperforms all other fully or linearly entangled circuits in addition to classical algorithms for most of the features. In fact, as the feature increases the efficiency of our proposed fully entangled model also increases significantly.  

\end{abstract}
\maketitle
\section{Introduction}
Machine learning is a branch of artificial intelligence that efficiently uses the underlying data to generate insightful classifications and predictions with the help of algorithms- while gradually improving the accuracy using historical data or information \cite{michalski2013machine}. Machine learning algorithms evolve from mathematical models reinforcing the prognosis associated with the dataset which in-turn further facilitates the informed decision process. Such algorithms help a machine to work efficiently even for analysing massive quantity of data. With the impact of machine learning algorithms on science and technology, it is imperative to develop and analyse new machine learning algorithms. In fact, the last two decades have witnessed an enormous transformation in terms of algorithms and the area has become a leading research area across academia and industry. Another area that has impacted academic and industrial progress is quantum information and computation. Quantum algorithms- based on the fundamental laws of quantum mechanics- have shown optimal assurance to offer efficient applications to material science \cite{barkoutsos2021quantum,bauer2020quantum}, computational chemistry \cite{armaos2020computational}, cryptography \cite{gisin2002quantum,bennett1992experimental}, traffic navigation \cite{yarkoni2020quantum,neukart2017traffic}, security \cite{chong2010quantum,Pirandola:20}, finance \cite{rebentrost2018quantum}, biology \cite{fedorov2021towards} and machine learning \cite{biamonte2017quantum}. Both, machine learning and quantum computing, deal with large-scale networks to effectively address critical problems, therefore, a natural question is to analyse the potentials of quantum computing in the realms of machine learning, i.e., quantum machine learning. Quantum computing, in comparison to classical computation, is believed to assist machine learning algorithms in enumerating distance or similarity between data-points with an adequate speedup \cite{aimeur2007quantum,wiebe2015quantum}. The advantage can be translated in terms of time and space complexity or can be obtained using  adiabatic quantum computing for attaining optimum results \cite{pudenz2013quantum}. Moreover, with the role of entanglement and non-classical correlations being established as the reasons for several potential applications in quantum information and computation \cite{zeilinger1998quantum,kempe1999multiparticle}, quantum machine learning algorithms are conceived to be much more powerful and effective as compared to their classical counterparts. \par
In addition to the quantum machine learning algorithms, data is also an important factor to achieve the desired accuracy \cite{budach2022effects,gaafar2022comparative}. Among various types of data, text data plays a key role in daily life, e.g., placing an online order, booking tickets or buying food online, reviews of movies or stores etc. With the advent of e-commerce, one of the most desirable requirements is the availability of reviews or compact surveys for every product before placing an order. Therefore, analysing the available data as positive or negative without any error or biasedness is extremely important and challenging. The classification or analysis of textual data is therefore known as sentiment analysis or opinion mining. Sentiment analysis not only helps users to classify the opinion but also the industry to improve as per the market requirements. In this article, we revisit sentiment analysis by proposing new quantum circuits to classify the data and to compare the efficiencies of our circuits with classical machine learning algorithms in addition to quantum circuits as proposed in \cite{havlivcek2019supervised,huang2021power}.   

\subsection{Motivation}
Machine learning algorithms are  historically considered to be effective in classifying text data as positive or negative sentiments.  Whether a statement is positive or negative can be effectively predicted by examining correlations among words which can be better analyzed using entanglement and non-local correlations that are  fundamental features of quantum computing.  Therefore, quantum machine learning algorithms have been developed for efficient classification of data- these algorithms have further shown modest expressibility compared to classical algorithms \cite{saini2020quantum,shaydulin2022importance}. Although quantum machine learning algorithms have shown a promising path to efficiently classify data, recent quantum kernel algorithms have  shown a decreasing trend in accuracy as the number of qubits increase \cite{huang2021power,shaydulin2022importance}. In view of the available noisy intermediate-scale quantum computers, the performance of quantum kernel algorithms depends on feature map circuits, optimization and accuracy of circuit depth. Here, we have further analysed another factor, i.e. correlation between features as an important hyperparameter for studying efficiency of quantum kernels. Analogous to classical machine learning algorithms, these variables or hyperparameters need to be tuned for realizing a quantum advantage. \par
In this paper, we readdress the problem of classifying data using our proposed quantum circuits considering circuit depth and entanglement between features. For analysing the efficiency of quantum kernels, we use linear as well as fully entangled circuits. Interestingly, our results show a much better accuracy, precision, recall and F1-score in comparison to classical algorithms including classical support vector machine and quantum circuits proposed in \cite{havlivcek2019supervised,huang2021power}. In order to compare the efficacy of our results, we also modify circuits proposed in \cite{havlivcek2019supervised,huang2021power} using linear and full entanglement. We show that our fully entangled circuit succeeds in outperforming quantum circuits proposed in \cite{havlivcek2019supervised,huang2021power} for different evaluation metrics. Moreover, our fully entangled circuit also results in accomplishing better efficiency for classification in comparison to the classical support vector machine. Surprisingly, with increasing number of features, our proposed circuit leads to significant increase in efficiency in comparison to all classical and quantum models analyzed here. The modified fully entangled circuits using \cite{havlivcek2019supervised,huang2021power} show better efficacy in comparison to the modified linearly entangled circuits using \cite{havlivcek2019supervised,huang2021power} for maximum number of features, however in both cases, accuracy and precision are less than the one obtained using classical support vector machine. \par
In addition to the IMDb review data, we also validate our circuits using the benchmarked Iris data \cite{Dua:2019} confirming the proposed fully entangled circuit as the most efficient circuit in comparison to other linearly or fully entangled circuits and Classical SVM.

\section{Preliminary and Fundamentals}
Sentiments are opinions, thoughts, or feelings of an author towards an object, an entity, or an event. Sentiment analysis or document level sentiment classification-  used in social media monitoring, recommendation system, customer service, brand monitoring, stock market prediction, product analysis to name a few \cite{vanaja2018aspect,chen2016exploring,isah2014social,abbasi2021tourism}- classifies a sentence by considering that sentence as a single information unit which includes positive, negative, or neutral statements \cite{liu2010sentiment}. Sentiment analysis, also known as text classification, uses well established supervised learning models such as support vector machine (SVM), gradient boosting classifier (GBC) and decision tree (DT) \cite{ye2009sentiment, pang2002thumbs, go2009twitter, kalaivani2013sentiment, athanasiou2017novel}. \par
\begin{figure}[!htb]
    \centering
    \includegraphics[width=0.8\linewidth]{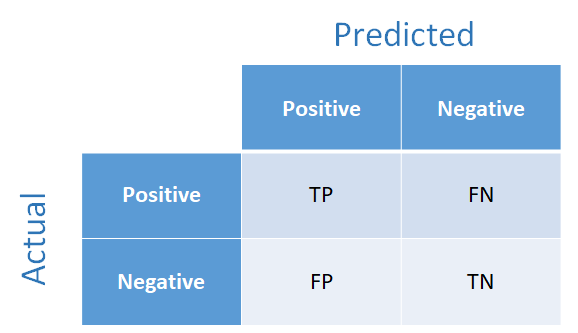}
    \caption{Confusion matrix for performance evaluation for a problem involving two classes}
    \label{Figure:confusion}
\end{figure}

Evaluating a classification model is an essential step to measure its performance using an unknown data. In general, the performance of the classification model is evaluated in terms of accuracy, precision, recall, and F1- score. The fundamental characteristic of these evaluation metrics is the confusion matrix which can be used to record the classification and misclassification. Fig. \ref{Figure:confusion} demonstrates a confusion matrix containing correct classifications and misclassifications in the form of four important values, i.e., TP - True Positive, FP - False Positive, FN - False Negative and TN - True Negative \cite{hicks2022evaluation}. 
Mathematically, different evaluation metrics are represented as 
\begin{equation}
    Accuracy = \frac{TP+TN}{TP+FP+FN+TN}
\end{equation}
\begin{eqnarray}
    Precision(P) = \frac{TP}{TP+FP} \nonumber \\
      Precision(N) = \frac{TN}{TN+FN}
\end{eqnarray}
\begin{eqnarray}
    Recall(P) = \frac{TP}{TP + FN} \nonumber \\
    Recall(N) = \frac{TN}{TN + FP}
\end{eqnarray}
\begin{equation}
    F1-Score = 2*\left(\frac{Precision * Recall}{Precision + Recall}\right)
\end{equation}
\begin{eqnarray}
    True Positive Rate (TPR) = \frac{TP}{TP + FN} \nonumber \\
    False Positive Rate (FPR) = \frac{FP}{TN + FP}
\end{eqnarray}
In the next subsection, we briefly describe fundamental concepts of quantum computing and quantum machine learning to facilitate the discussions on results obtained in this article. 

\subsection{Quantum Computing and Quantum Machine Learning Algorithms}
The statistical nature of quantum theory puzzled many physicist including Einstein which further prompted him and couple of his colleagues to raise questions on foundations of quantum mechanics, now known as Einstein-Podolsky-Rosen (EPR) paradox \cite{einstein1935can}. However, J. S. Bell proposed another thought experiment by designing an inequality - the Bell inequality - whose violation confirms the presence of correlations between subsystems of a composite system contradicting the very assumptions of locality and realism as advocated by EPR \cite{bell1964einstein}. The fundamental debate has now been translated to the practical implementation of a fault tolerant quantum computer which can utilize nonlocal aspects of quantum theory to perform efficient tasks which are either not possible using classical resources or may require exponential time and memory. The technological developments have ensured a rapid advancements towards realization of a quantum computer, at least with more than 100 qubits. Even with intricate challenges of controlling, measuring, and accessing quantum information, the future of quantum computation to address large-scale computational issues is highly promising. Quantum computing has the capability to embrace machine learning by strengthening the analysis through quantum machine learning algorithms. In this subsection, we briefly discuss foundational aspects of quantum computing and quantum machine learning algorithms.
\subsubsection{Quantum Computing}
Quantum computation uses fundamental aspects of quantum theory for the paradigm shift in computing from classical to quantum regime \cite{nielsen2002quantum}. For a two-level system, the fundamental unit is termed as a qubit which can be represented as an arbitrary linear superposition of two orthogonal basis vectors. To be precise,  a qubit- using computational basis states $\ket{0}$ and $\ket{1}$- can be described as  $\ket{\psi} = \alpha\ket{0}+\beta\ket{1}$ where $\alpha$, $\beta$ $ \in $ $\mathbb{C}$ and $\abs{\alpha}^2 + \abs{\beta}^2 = 1$. For multiple qubits, the inherent nature of superposition leads to another interesting quantum mechanical phenomena known as entanglement \cite{plenio2014introduction}. For two qubit pure entangled systems, the four maximally entangled Bell states are extensively used for optimal communication and computational tasks \cite{bennett1993teleporting,bennett1992communication,shimizu1999communication,shukla2014protocols}. For more than two qubit systems, there exist multiple classes of entangled states due to the complex nature of multiqubit entanglement \cite{dur2000three,sabin2008classification}. The entangled systems violating the Bell or Bell-type inequalities exhibit nonlocal correlations with no classical analogues. The nonclassical correlations are not limited to entangled systems only, but can be extended to separable ones as well \cite{knill1998power,datta2008quantum}. It has been well established now that entanglement and nonclassical correlations offer efficient advantages to quantum computation over classical resources \cite{meyer2000sophisticated,kenigsberg2006quantum,singh2018analysing,kaur2020nonlocality,faujdar2021comparative}. \par
For circuit based implementations- analogous to classical circuits- quantum circuits also comprise of wires and quantum gates. However, the only constraint quantum gates need to satisfy is that the operator representing the gate must be a unitary operator- unlike classical gates quantum gates are reversible \cite{nielsen2002quantum, williams2011quantum}. 
\subsubsection{Quantum Support Vector Machine (QSVM)}
Supervised quantum models are summarised as kernel methods where classification of the data is performed in a high dimensional Hilbert space accessible through inner products \cite{schuld2021supervised}. Quantum support vector machine represents a quantum-classical hybrid approach where kernel value is computed using a quantum computer and the classification process is performed using a classical computer. For computing kernel values, classical data  must be converted into quantum states. Therefore, first the classical data points are transformed to quantum states using feature mapping techniques, and then the kernel values are estimated. In fact, the kernel is analogous to calculating inner products between quantum states \cite{havlivcek2019supervised} such that
\begin{equation}
K(\Vec{x_{i}},\Vec{x_{j}})=\abs{\bra{\psi_{\Vec{x_i}}}\ket{\psi_{\Vec{x_j}}}}^2
\label{qsvm}
\end{equation}
where $\Vec{x_i}, \Vec{x_j}$ $\in$ dataset and $\abs{\bra{\psi_{\Vec{x_i}}}\ket{\psi_{\Vec{x_j}}}}^2$ can be further defined as
\begin{equation}
\abs{\bra{\psi_{\Vec{x_i}}}\ket{\psi_{\Vec{x_j}}}}^2 = 
\abs{\bra{0^n}\mathcal{U}^{\dagger}_{\psi(\Vec{x_i})}\mathcal{U}_{\psi({\Vec{x_j}})}\ket{0^n}}^2
\label{qsvm2}
\end{equation}
From Eq. \eqref{qsvm2}, one clearly see that kernel values can be practically computed using quantum circuits by first evolving the initial state $\ket{0}$ under the influence of a unitary operator $\mathcal{U}_{\psi(x)}$ and then finally measuring the state in computational basis. Eventually, the computed kernel value is passed to the Classical SVM for classification, i.e., 
\begin{equation}
    class = sign \left(\sum_{i} y_{i}\alpha_{i}K(\Vec{x_{i}},\Vec{z})+b\right)
\end{equation}
where $b$ represents offset. Here, if the sign of Classical SVM is positive then the state belongs to a positive class and if the sign of Classical SVM is negative then the state belongs to a negative class \cite{cortes1995support}. In this work, we have propose new quantum circuits to compute kernel values and compare the performance of proposed circuits with classical machine learning algorithms and quantum circuits in \cite{havlivcek2019supervised,huang2021power}. The composition of proposed circuits will be discussed in Sec. \ref{feature_mapping}.

\begin{figure*}
    \centering
    \includegraphics[height=6cm,width=6in]{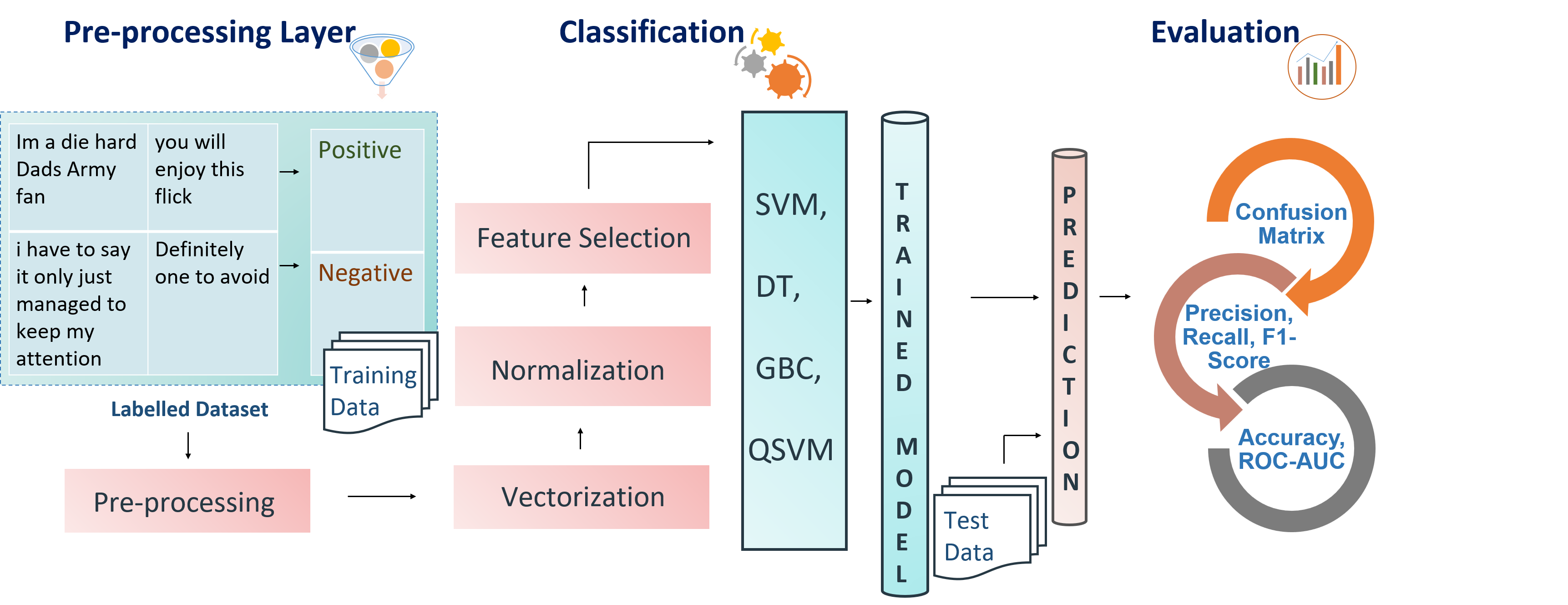}
    \caption{The proposed pipeline for analysis of the text data}
    \label{fig:pipeline}
\end{figure*}
\section{Related Work}
The fusion of machine learning with quantum computing promises several new and exciting algorithms- including the variational  algorithms- for optimal speed-up to large-scale data driven tasks \cite{benedetti2019parameterized,biamonte2017quantum,de2022survey}. In this section, we briefly discuss some of the algorithms developed in the field of quantum machine learning. \par
Quantum computing benefits classical supervised and unsupervised machine learning algorithms. Rebentrost \textit{et al.}  have shown quantum version of the support vector machine using phase estimation and quantum matrix inversion algorithm for maximizing the speed \cite{rebentrost2014quantum}. Another significant quantum support vector machine was proposed by Havelick \textit{et al.} which was implemented on IBMQ open-source software \cite{havlivcek2019supervised}. The proposal was based on evaluating the kernel of classical support vector machine quantum mechanically,  analogous to computing the inner product of quantum states. Lu \textit{et al.} \cite{lu2014quantum} demonstrated the construction of a quantum decision tree where the von-Neumann entropy determines the split over the node and fidelity measures the distance between two quantum states. On the experimental front, the efficient use of quantum computing in classifying high-dimensional vectors and mathematical routines was further established using an entanglement based classification \cite{cai2015entanglement}. Schuld \textit{et al.} introduced a quantum pattern classification algorithm based on k-nearest neighbours where the distance between states is measured using the Hamming distance  \cite{schuld2014quantum}. Moreover, they also utilized the proposed algorithm for classifying Modified National Institute of Standards and Technology (MNIST) dataset \cite{schuld2014quantum}. The results facilitated Ruan \textit{et al.} to design a quantum k-nearest neighbour algorithm for calculating the Hamming distance between training inputs and the testing state \cite{ruan2017quantum}. On similar lines, Wiebe \textit{et al.}  also presented a quantum algorithm for nearest neighbour and k-means clustering algorithm, evolving quantum computing into the realms of  unsupervised machine learning \cite{wiebe2014quantum2}. For this, they have proposed two methods for calculating the distance between vectors using a  quantum computer, namely, inner product and euclidean method. Another hybrid k-means clustering algorithm was proposed by Sarma \textit{et al.} for finding k-subsets having less dissimilarity among its members  \cite{sarma2019quantum}. In this case, euclidean distance was evaluated using quantum circuits and classification of  data points was  executed classically. They further implemented their algorithm on the IBMQ open-source software and demonstrated its accuracy to be  better then the classical k-means algorithm. There are several other instances of significant contributions to quantum machine learning algorithms for supervised as well as unsupervised learning in comparison to classical algorithms \cite{aimeur2013quantum,aimeur2006machine,aimeur2007quantum,kerenidis2019q,otterbach2017unsupervised,alvarez2017supervised,kyriienko2022unsupervised}.\par
The first instance of understanding the neural networking of human brain using quantum computing was demonstrated by Kak \cite{kak1995quantum}. Menneer and Narayanan further developed the idea to discuss a practical approach of using quantum computing in neural networks \cite{menneer1995quantum}. In order to obtain the quantum inspired neural network, they first trained  a number of neural networks and then evaluated the superposition of these networks. Ventura and Martinez further proposed a method for quantum associative memory based on Grover's algorithm \cite{grover1998quantum} and demonstrated an exponential storage capacity of associative memory \cite{ventura2000quantum}. Based on the results of Ventura and Martinez, Zhou \textit{et al.} developed a model for the quantum associative neural network to generate a quantum binary decision for quantum array to store patterns \cite{zhou2012quantum}. Further, in order to improve the performance of neural networks and conventional neural networks, quantum-inspired neurons, and activation functions have also been designed \cite{konar2022optimized,shi2022quantum}. These improved versions of different deep learning algorithms can be useful in crucial fields like medical imaging. \par
\begin{figure*}
    \centering
    \includegraphics[height=7cm,width=6in]{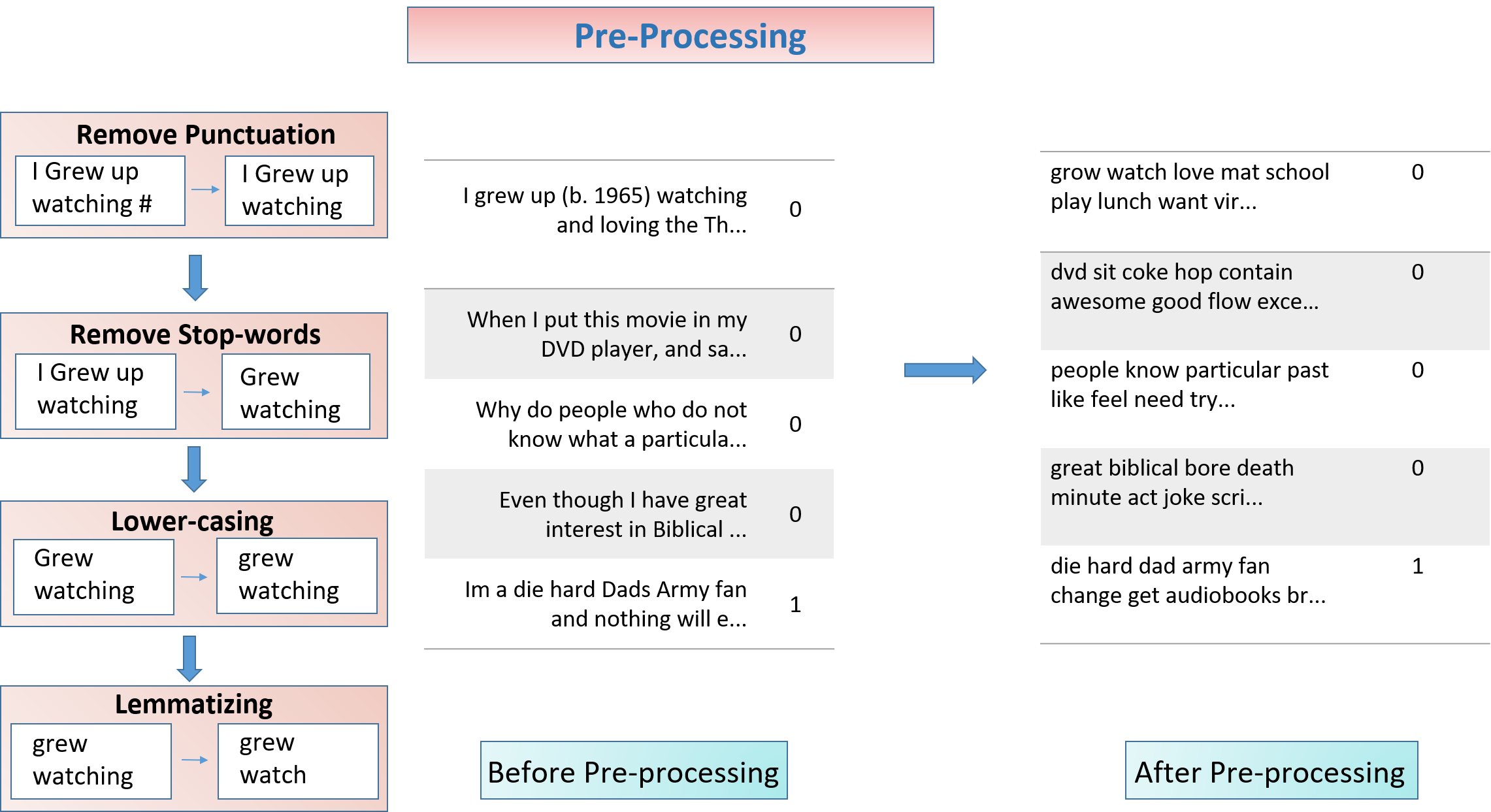}
    \caption{Pre-processing steps for useful feature extraction from the raw dataset}
    \label{fig:pre-processing}
\end{figure*}
Evidently, in all quantum and classical algorithms, dataset plays a crucial role. The performance of algorithms always depend on the type of data points, dimension of the dataset, outliers etc. In view of this, Huang \textit{et al.} \cite{huang2021power} further established the power of data by considering  different quantum kernels, and advantages of quantum algorithms in comparison to classical algorithms  when the geometric difference is very high. In order to capture the distinct sentiment information, Zhang \textit{et al.} proposed a quantum-inspired sentiment representation model \cite{zhang2019quantum}. This model extracts sentiment phrases and represents sentiment information rooted in phrases as a collection of projectors. Considering feature selection to play  a key role in improving the overall accuracy of machine learning models, Chakraborthy \textit{et al.} depicted a quantum mechanical  feature selection algorithm using a graph-theoretic approach \cite{chakraborty2020hybrid}. They further demonstrated that their feature selection strategy outperformed many of the classical feature selection strategies. Therefore, quantum-inspired classical machine learning algorithms are considered as an efficient approach to increase the efficiency and accuracy of such algorithms.  However, a very few instances can be found related to classification and clustering of real-time classical data using quantum machine learning algorithms. Clearly, the performance of an algorithm can be analysed by testing it over different types of real-valued datasets. The quantum support vector machine was originally tested over the ad-hoc numerical data and the sentiment analysis of an unseen data using quantum algorithms is still at a nascent stage.  Therefore, in this study, we perform a comparative sentiment analysis using quantum and classical algorithms based on different performance metrics. Depending on the number of features, our analysis quantifies the advantages of quantum machine learning algorithms for classifying the IMDb review data. 

\section{BASIC STRUCTURE OF THE PROPOSED APPROACH}

In the proposed approach, as depicted in Fig. \ref{fig:pipeline}, we analyse sentiments of textual data considering its applications in spam filtering, intention mining, product analysis, and market research etc. In this work, we analyse sentiments of textual data considering its applications in spam filtering, intention mining, product analysis and market research, to name a few. For this, we use the  IMDb movie review dataset downloaded from Kaggle \cite{dataset} in which reviews are stored in text format with 40000 reviews and a single review containing 200 or more words. As the raw text data generally includes redundant information with no adequate contribution to the classification process, Fig. \ref{fig:pre-processing} shows all pre-processing steps use in this work and their effects on the text data. \par 
In order to convert reviews into numerical form, the next important step is to perform vectorization of the pre-processed text data. In the proposed strategy, we use CountVectorizer, also known as Bag of Words (BoW). For this, BoW creates a set of words present in the whole dataset and computes the frequency for each token by analyzing the occurrence of words in a document. BoW produces a sparse matrix where some of the tokens acquire a large value and others do not. Hence, we further normalize the resulting dataset by dividing feature values with the square root of the sum square of all  features, given as
\begin{equation}
    f^{'}_{j} = \frac{f_{j}}{\sqrt{\sum_{j=1}^{d} f_{j}^{2}}}
\end{equation}
Here, $f_{j}$ is the feature value, and $d$ represents dimensions of the data.
\subsection{Feature Selection}
For a better analysis of the text data, we specifically require features which are significant during classification to avoid over-fitting. For feature selection, we first use the recursive feature elimination (RFE) method which recursively eliminates features not contributing to the classification. As RFE requires an estimator for assigning weights to features, we employ Random Forest \cite{jeon2020hybrid, breiman2001random} and further characterize these features for their significance in classification using   LASSO (Least Absolute Shrinkage and Selection Operator) which regularizes features by penalizing \cite{tibshirani1996regression,ghosh2005classification}. For this, we use the normalized data $(x_i,y_i)$ to approximately compute $y_i$  while minimizing the cost function $C(w)$. As a key point, LASSO adds the sum of absolute value of a coefficient parameter to the cost function given as
\begin{equation}
  w^{lasso} = argmin[C(w) + \sum_{j=1}^{d} \lambda \abs{w_{j}}]
\end{equation}
where
\begin{equation}
    C(w) = \sum_{i=1}^{n}\left(y_{i}-\sum_{j=1}^{d}w_{j}x_{ij}\right)^2
\end{equation}
Here $C(w)$ is the cost function and $w$ is the coefficient parameter which is determined by $\lambda$. Therefore, $\lambda$ is further optimized to attain best coefficient value corresponding to essential features. \par
\begin{figure}
    \centering
    \includegraphics[width=0.48\textwidth]{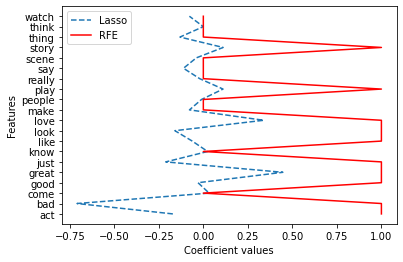}
    \caption{Analyzing significant features based on coefficient values using RFE and LASSO}
    \label{fig:feature_sel}
\end{figure}

Fig. \ref{fig:feature_sel} clearly indicates that RFE shows features to be either significant or not by analyzing 1 or 0 value corresponding to features; whereas LASSO shows the coefficient value for features. The figure further demonstrates that features attaining a 0 value for RFE also get a coefficient value nearly equal to 0 and therefore get the last preference while selecting features. \\
Finally, we construct the database using selected features and perform our analysis using classical and quantum machine learning algorithms. In contrast to classical machine learning algorithms, for the quantum machine learning algorithms, we first construct a model for mapping classical features into quantum states. Therefore, we now proceed to discuss the model for mapping classical data to quantum states in the following subsection. 
\subsection{Feature Mapping}\label{feature_mapping}
\begin{figure}
    \centering
    \includegraphics[width=0.48\textwidth]{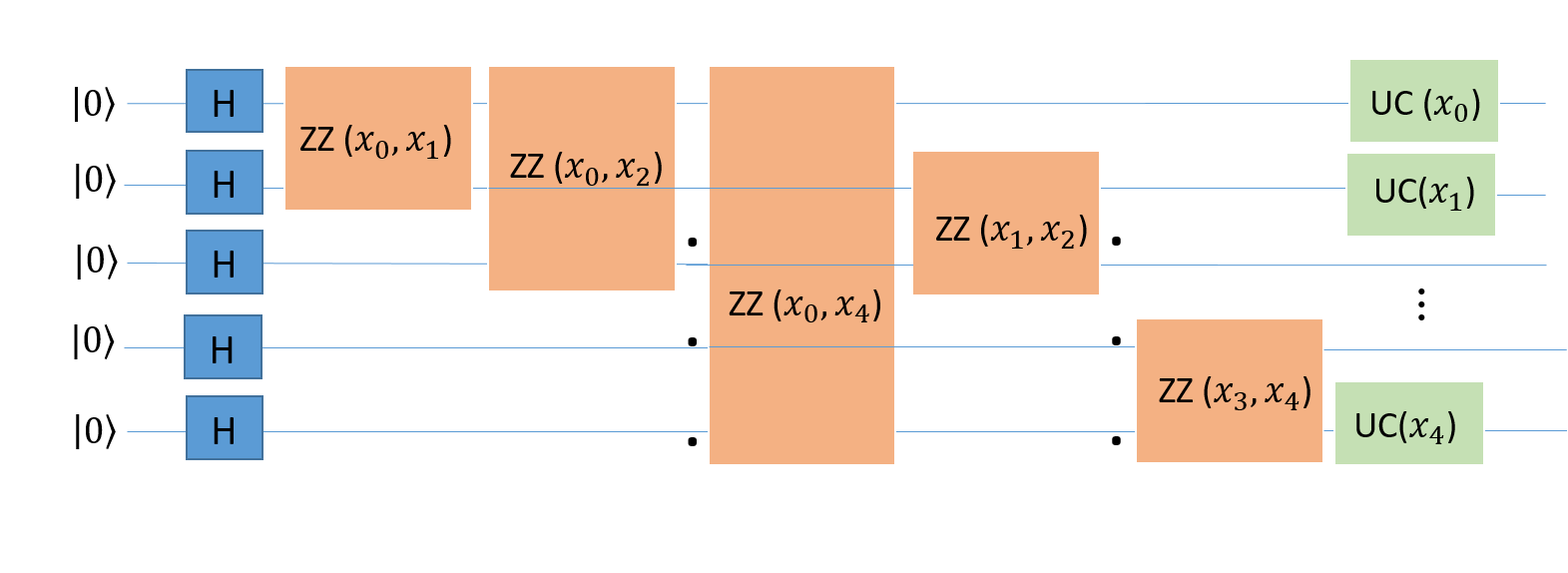}
    \caption{The feature map circuit used for fully entangled proposed circuit to embed classical data into the quantum state}
    \label{fig:feature_map}
\end{figure}

Once the features are selected as per requirements of the near-term quantum computer, we proceed with the development of a feature mapping model. For feature mapping, the classical data $\Vec{x_{i}}$ $\in$ $\mathbb{R}^n$ is embedded into quantum states $\ket{\phi(\Vec{x_{i}})}$ through the unitary transformation $\mathcal{U}_{\phi(\Vec{x})}$ such that
\begin{equation}\label{FE}
    \ket{\phi(\Vec{x_{i}})} = \bigotimes_{j=1}^{d} \mathcal{U}_{\phi(\Vec{x})}(x_j)
\end{equation}
\begin{figure}
    \centering
    \includegraphics[width=0.48\textwidth]{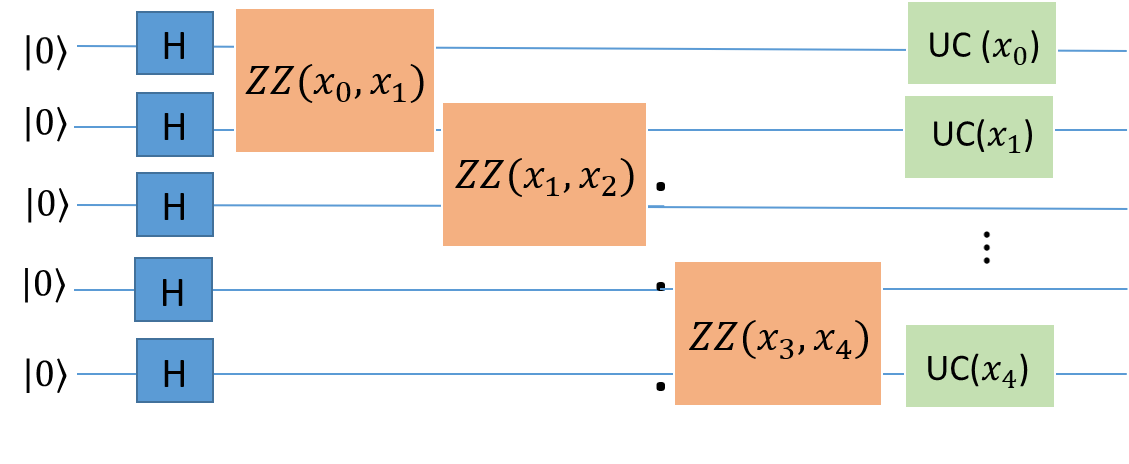}
    \caption{The feature map circuit used for linearly entangled proposed circuit to embed classical data into the quantum state}
    \label{fig:ZZ_Y_linear}
\end{figure}
Fig. \ref{fig:feature_map} represents the proposed circuit implementation for the unitary transformation. We initiate the circuit with $\ket{0}$ states and create a superposition state by applying Hadamard gates on all qubits. The number of features determines the number of qubits used in the circuit. The superposition state is then evolved as shown in Fig. \ref{fig:ZZ_Y_linear}. Here ZZ gate is a combination of CNOTs and rotation gates along Z axis such that the angle is determined by a function given as 
\begin{equation}
    \phi(x_j,x_{j^{'}}) = (\pi-x_j)(\pi-x_{j^{'}})
\end{equation}
where $\phi(x_j,x_{j^{'}})$ is a classical non-linear function which depends on feature values $x_j$ and $x_{j^{'}}$ of a document associated with the dataset. The CNOT gates entangle a qubit with its successor. In Fig. \ref{fig:feature_map}, UC gate is a  combination of $R_{x}$ and $R_y$ gates, where $R_x$ is a rotation gate with a fixed angle and $R_y$ is a feature value $(x_j)$ dependent gate. Using the circuit, Eq. \ref{FE} can be re-written as
\begin{equation}
   \ket{\phi(\Vec{x_{i}})} = \mathcal{U}_{\phi(\Vec{x_{i}})} \bigotimes_{j=0}^{d-1}  H ^{j} \ket{0^{j}}
\end{equation}
where $\mathcal{U}_{\phi(\Vec{x_i})}$ is
\begin{equation}
    \mathcal{U}_{\phi(\Vec{x_i})} = exp \left(\sum_{j=0}^{d-1} \sum_{j^{'}=j+1}^{d-1}  Z_{j}Z_{j^{'}} \phi(x_j,x_{j^{'}}) + \sum_{j=0}^{d-1} UC_{j}(x_{j})\right)
\end{equation}
We further consider  two more feature map circuits inspired by instantaneous quantum polynomial (IQP) and Hamiltonian evolution circuit (HEC) or Hamiltonian circuit used in \cite{havlivcek2019supervised,huang2021power} where \cite{havlivcek2019supervised} shows that IQP circuit leads to comparable accuracy with classical support vector machine. The unitary operators for computing kernel values using IQP and Hamiltonian Circuit are given in Eqs. (15) and (16), respectively.
\begin{equation}
    \mathcal{U}_{\phi(\Vec{x_i})} = exp \left(\sum_{j=0}^{d-1} Z_{j}(x_{j}) + \sum_{j=0}^{d-1}\sum_{j^{'}=j+1}^{d-1} Z_{j}Z_{j^{'}}(\phi(x_j,x_{j^{'}})) \right)
\end{equation}
\begin{equation}
    \begin{aligned}
    \mathcal{U}_{\phi(\Vec{x_i})}  =
    exp \Bigg( \Bigg. \sum_{j=0}^{d-1}\sum_{j^{'}=j+1}^{d-1} (X_{j}X_{j^{'}} +  Y_{j} & Y_{j^{'}} +  Z_{j}Z_{j^{'}}) \\
    & (\phi(x_j,x_{j^{'}}))\Bigg. \Bigg)
    \end{aligned}
\end{equation}
Eventually, feature maps transform classical data points into higher dimensional space ($\mathbb{C}^{2n}$-dimensions) and facilitate computing the kernel value for the quantum support vector machine. 
\begin{figure}
    \centering
    \includegraphics[width=0.48\textwidth]{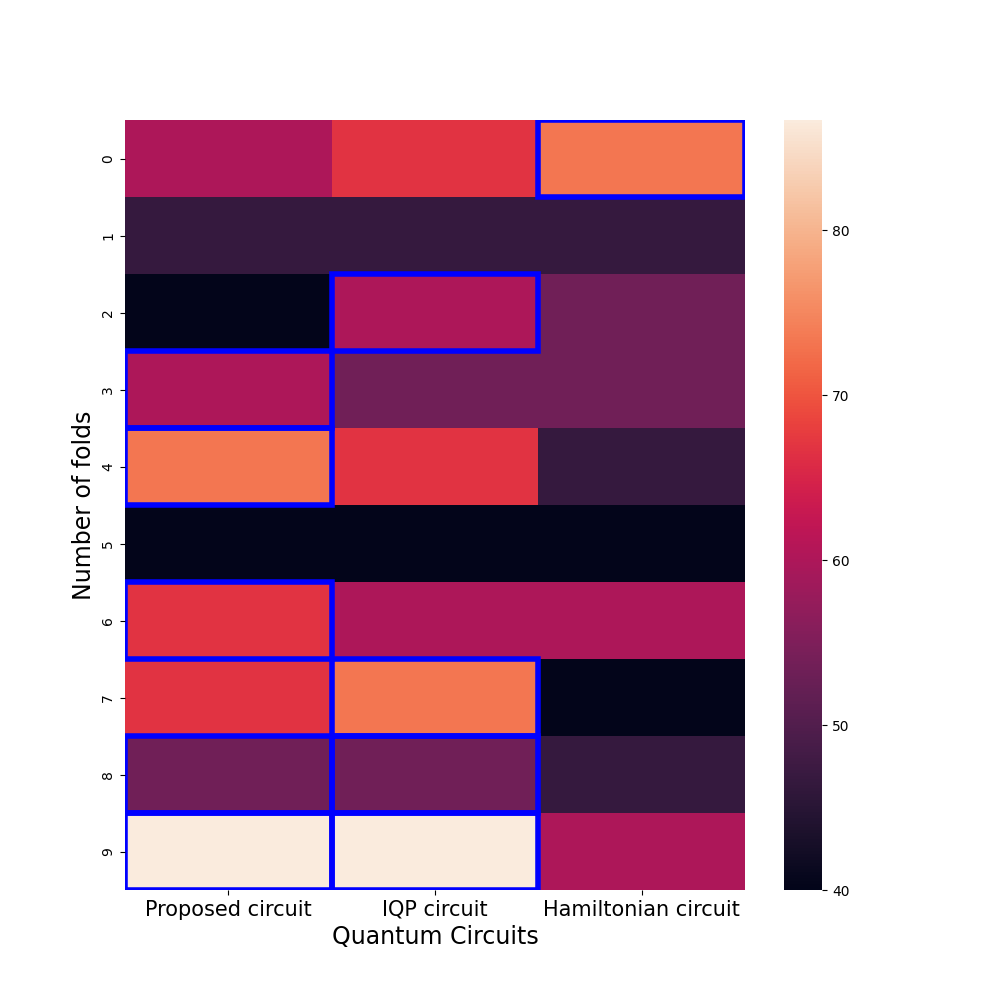}
    \caption{Prediction accuracy for Hamiltonian, IQP and proposed circuit with k-fold cross validation for linear entanglement}
    \label{fig:accuracy_linear}
\end{figure}
\begin{figure}
    \centering
    \includegraphics[width=0.48\textwidth]{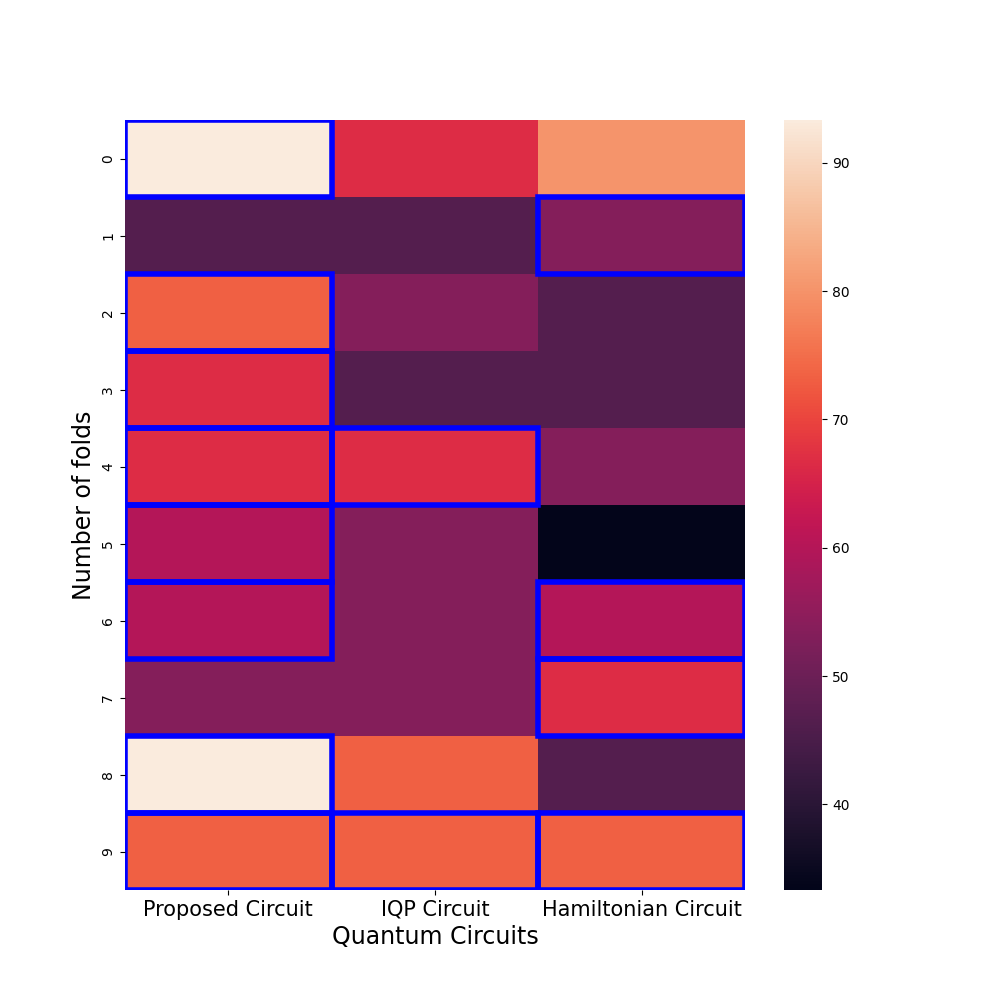}
    \caption{Prediction accuracy for Hamiltonian, IQP and proposed circuit with k-fold cross validation for full entanglement}
    \label{fig:accuracy_full}
\end{figure}
\section{Results}
We now proceed to observe effects of entanglement on performance of quantum circuits. For this, we have considered two different forms of entanglement- first is linear entanglement where each qubit is entangled only with its successive qubit, and second is full entanglement where every  qubit is entangled with all successive qubits. Figs. \ref{fig:ZZ_Y_linear}, and \ref{fig:feature_map} represent linear and full entanglement circuits, respectively. In addition, for evaluating efficiencies of these quantum kernel circuits, we first use a five-qubit IBM quantum computer. For efficient implementation of data over the available near-term quantum computer, we perform our computation with a random subset of the data with 150 samples. 
We further train and test the data multiple times using K-fold stratified cross-validation to develop a generalized model. For our purpose, we use 10-fold which samples the dataset 10-times into train and test data points.\par
Fig. \ref{fig:accuracy_linear} shows accuracy prediction for Hamiltonian, IQP and proposed circuit with qubits being linearly entangled. Similarly, Fig. \ref{fig:accuracy_full} demonstrates accuracy prediction as evaluated for Hamiltonian, IQP and proposed circuit with qubits being fully entangled. Further, for a comparative analysis, we also evaluate accuracy, precision, Recall and F1- score using Classical SVM, decision tree and gradient boosting classifier. Our results for linearly entangled circuits show that IQP circuit is more efficient in comparison to Hamiltonian and proposed circuit for all evaluation metrics. However, the Classical SVM exceeds in accuracy, precision and recall in comparison to all linearly entangled circuits and classical algorithms. We further find that the predictions of IQP linearly entangled circuit are very close to that of gradient boosting classifier. Interestingly, for the fully entangled circuits, the proposed circuit performs exceedingly well in comparison to all algorithms including Classical SVM and the other two quantum circuits for all evaluation metrics. The shift from linear to fully entangled circuits results in increase in efficiencies for the proposed and Hamiltonian circuits but surprisingly leads to the decreased efficiency of IQP circuit.\par
\begin{figure}
    \centering
    \includegraphics[width = 0.48\textwidth]{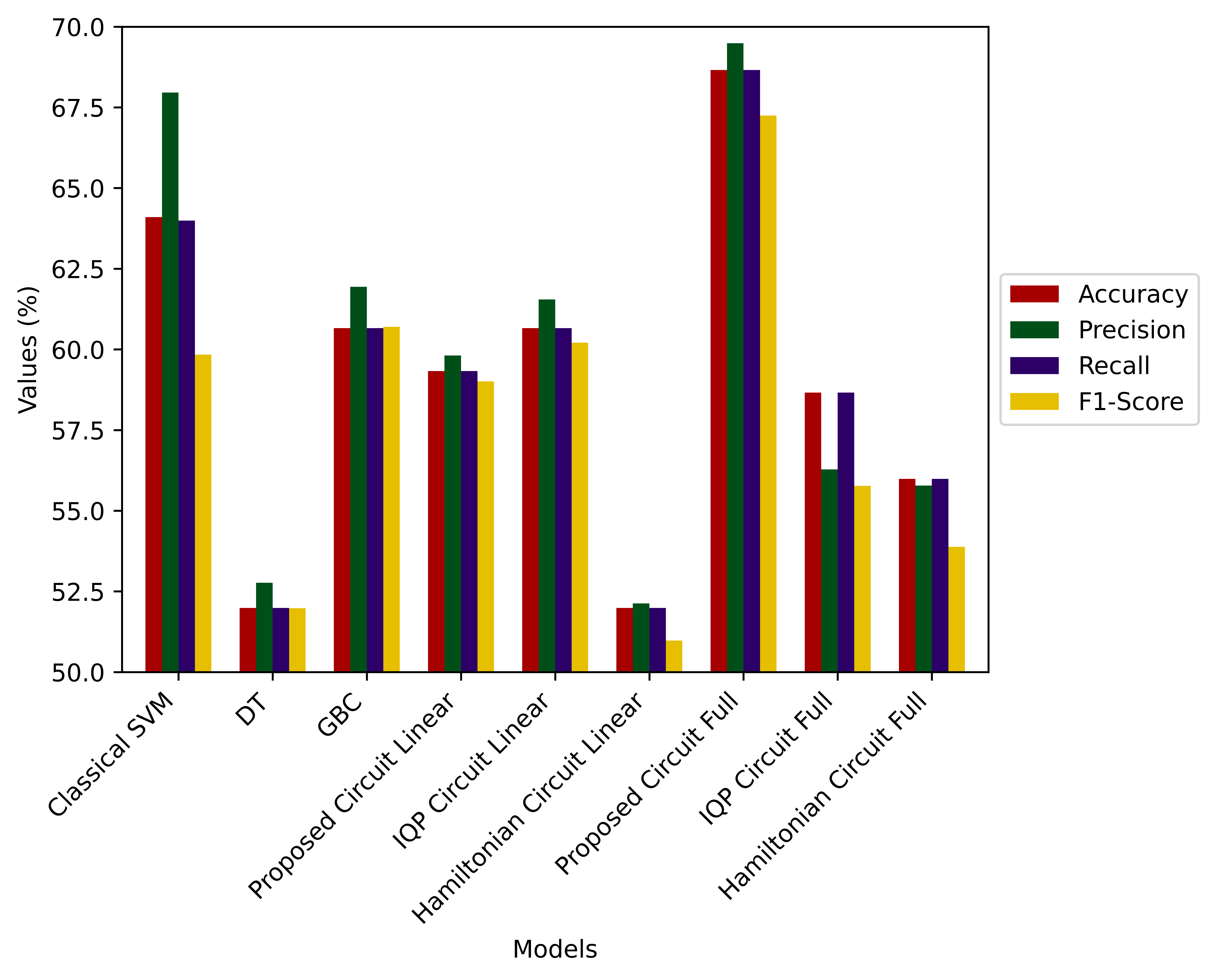}
    \caption{Performance measures of classical and quantum machine learning algorithms}
    \label{fig:aprf-graph}
\end{figure}

 
As discussed above, the performance of a classification model cannot be ascertained by analysing the accuracy only. The factors such as precision and recall further play a significant role in determining specificity and sensitivity of the model. Fig. \ref{fig:aprf-graph} demonstrates advantages of quantum support vector machine with full entanglement for evaluating efficiencies of algorithms in terms of accuracy, precision, recall and F1-score with 10-fold cross validation. Among classical models, Classical SVM is found to be the highest performing model in terms of accuracy, precision and recall which can be used as a benchmark for classical algorithms used in this article. \par
\begin{figure}[h]
    \centering
    \includegraphics[width = 0.48\textwidth]{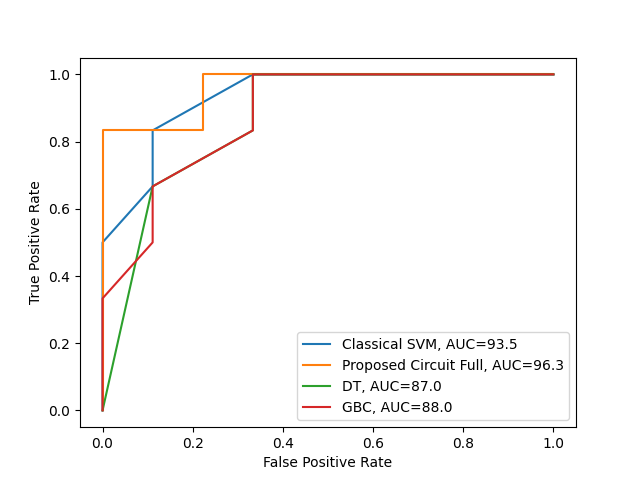}
    \caption{The ROC-curve between proposed fully entangled and classical model at their best ROC-AUC values}
    \label{fig:roc_curve}
\end{figure}

 
We also evaluate the performance of models using ROC-AUC values as depicted in Fig. \ref{fig:roc_curve} where AUC stands for area under ROC curve. The ROC curve represents a plot- for best ROC-AUC value from 10-fold runs- for true positive rate (TPR) versus false positive rate (FPR) where the TPR and FPR are predicted probabilities for test data points by respective models. Therefore, our results further suggest that the proposed fully entangled quantum circuit will predict any unseen instance as positive with the highest probability if that instance is actually positive. \par
\begin{figure}
    \centering
    \includegraphics[width=0.48\textwidth]{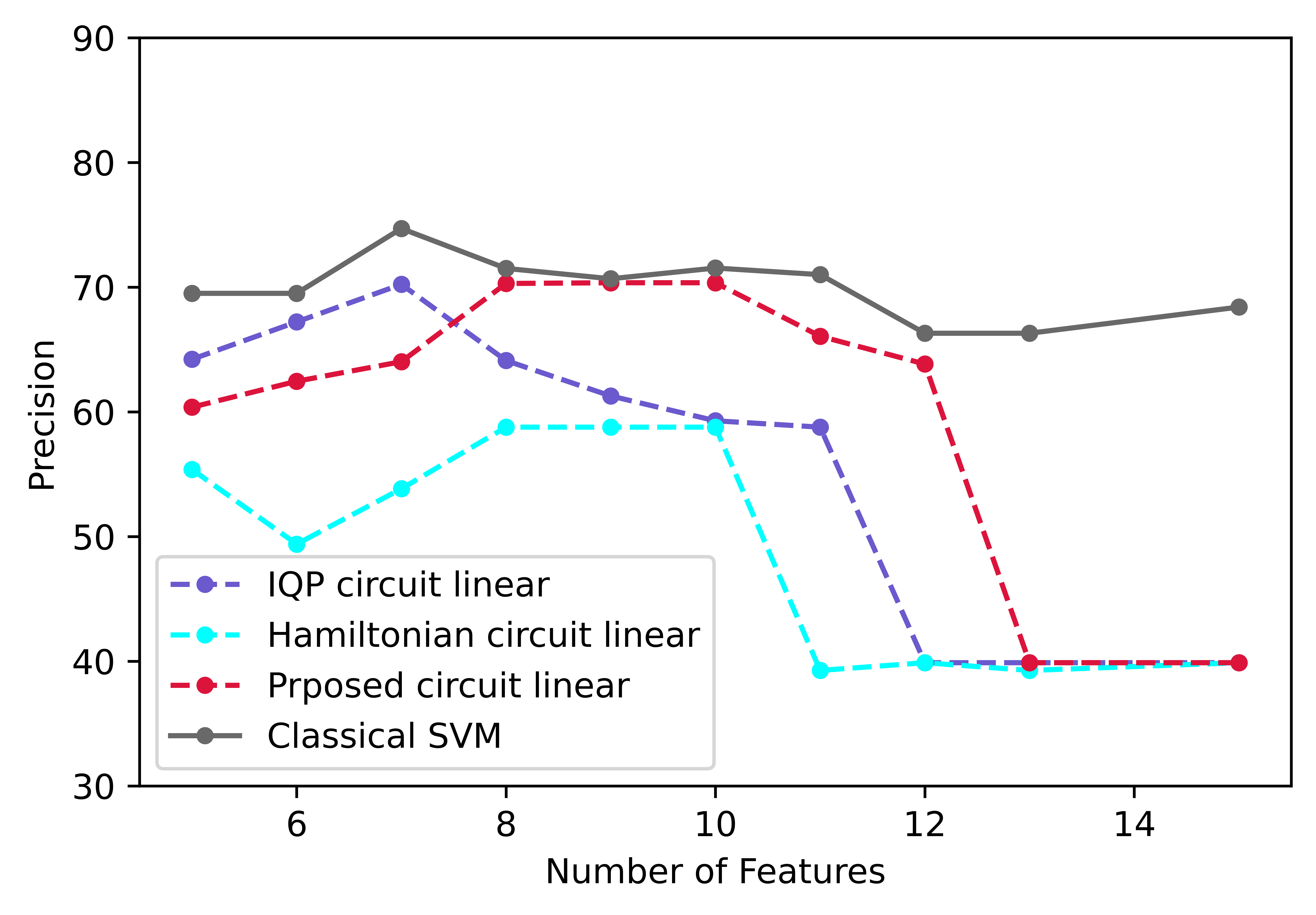}
    \caption{Precision score with increase number of features for all linear entangled circuits in comparison to Classical SVM}
    \label{fig:linear_ent}
\end{figure}
\begin{figure}
    \centering
    \includegraphics[width=0.48\textwidth]{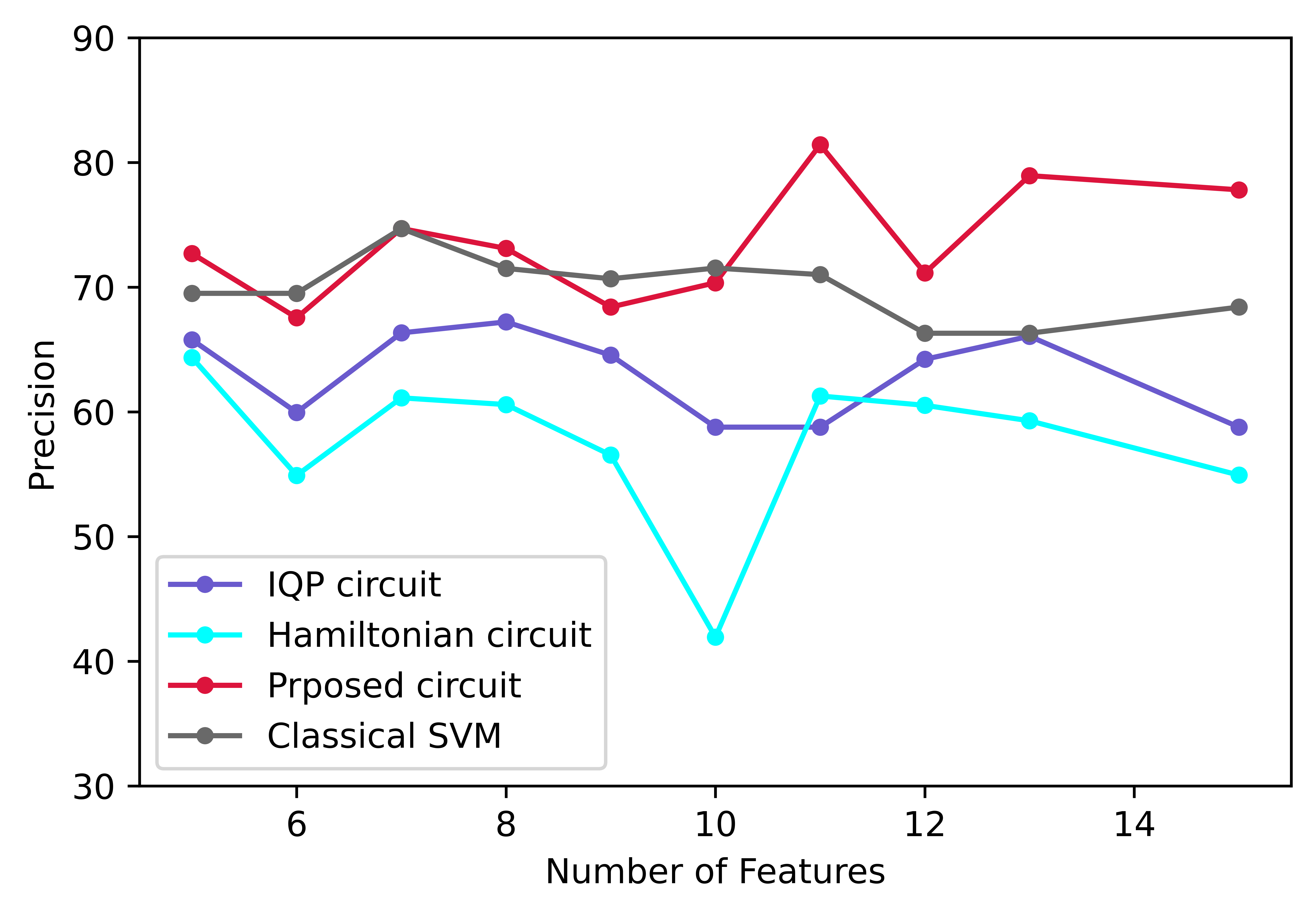}
    \caption{Precision score with increase number of features for all fully entangled circuits in comparison to Classical SVM}
    \label{fig:full_ent}
\end{figure}

 With the 10-fold cross-validation, we demonstrated the importance of fully entangled circuit as against the linearly entangled circuit and Classical SVM. We now proceed to increase number of features and to analyze our data using all algorithms with $25\%$ of data as test data and rest as train data. The purpose is to study and compare efficiencies of algorithms with increasing number of features.  Recently, Shaydulin and Wild \cite{shaydulin2022importance} and Huang \textit{et al.} \cite{huang2021power} have shown that efficiency of their respective quantum models decreases with increase in features. Here, we consider precision as the evaluation metric because of imbalanced positive and negative data points in data set. Figs. \ref{fig:linear_ent} and \ref{fig:full_ent} demonstrate precision score of linear and full entangled circuits, respectively with increased number of features and further compare their performances with classical support vector machine. Fig. \ref{fig:linear_ent} shows that the precision score of the proposed circuit is very close to the Classical SVM if the number of features are 8, 9, 10 or 12. However, Classical SVM performs the best in comparison to all linearly entangled circuits including the proposed circuit irrespective of number of features. \par
 Similar to the 10-fold cross validation, the proposed fully entangled circuit shows best precision score in most of the cases except when number of features are 6, 9 and 10 where the precision values are very close to the Classical SVM. As the number of features increase, the precision score for our proposed model is significantly more than any other quantum or classical algorithms. Surprisingly when we use 9 features, the linearly proposed circuit shows a better result than the proposed fully entangled circuit. Clearly, the highest precision score is obtained with 11 features for the proposed fully entangled circuit. Similar to the 10-fold cross validation, the fully entangled circuits offer a better perspective of analysis for precision score in comparison to the linearly entangled circuits. The precision score for Classical SVM exceeds the precision scores for linear and fully entangled IQP and Hamiltonian circuits except when the number of features are 13 where the precision score for Classical SVM and fully entangled IQP circuit are the same. Our results suggest that for linear entanglement, the efficiency of quantum circuits decrease with increasing number of features. However, such a linear relationship between efficiency and number of features could not be observed for fully entangled circuits. In fact, for the proposed fully entangled circuit, the efficiency increases with an increase in number of features. Therefore, our analysis suggests the importance of entanglement in determining the effectiveness of quantum kernel approaches and can potentially be used for a correlated dataset.\par
\begin{figure}
    \centering
    \includegraphics[width = 0.48\textwidth]{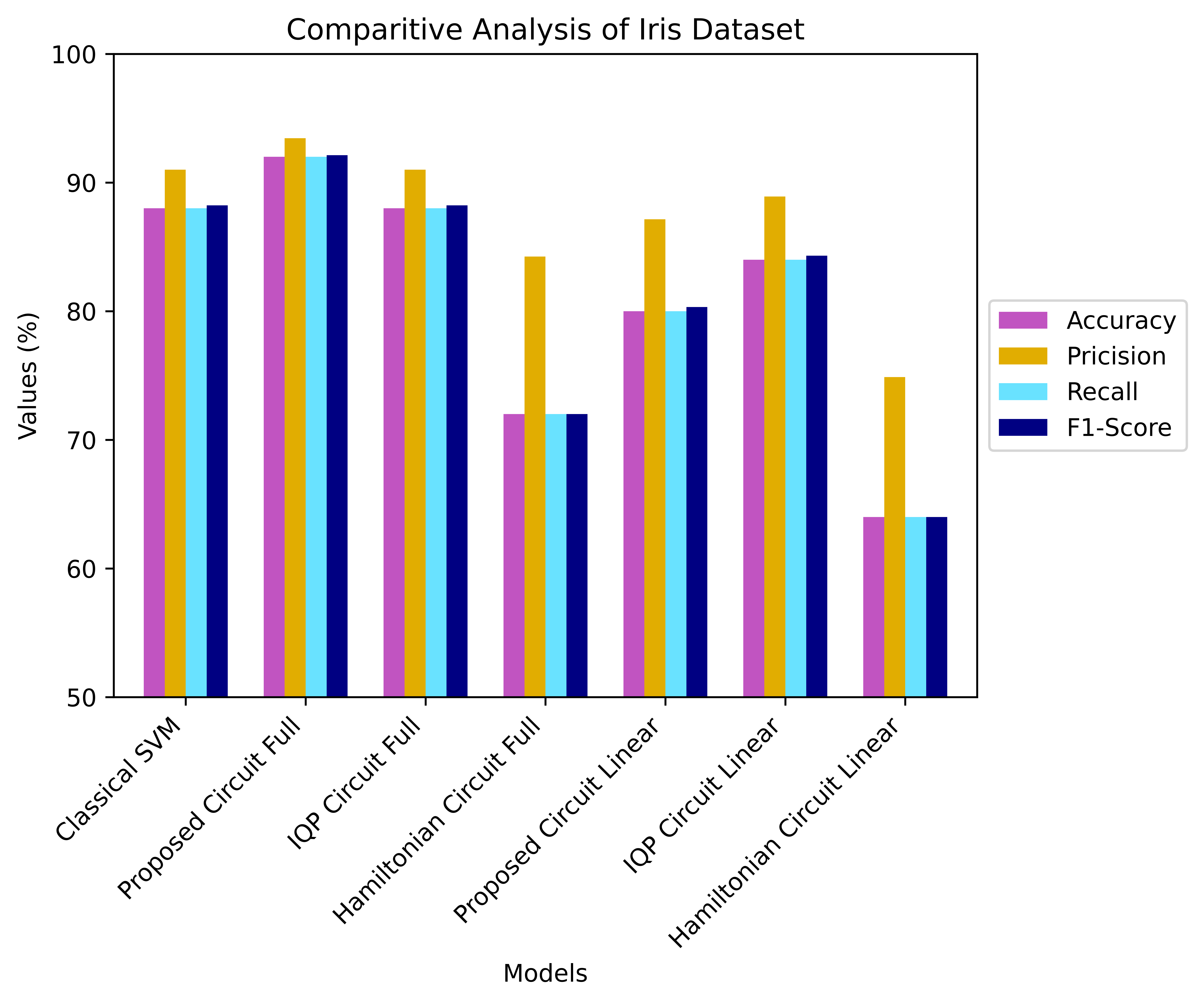}
    \caption{Performance measures of classical and quantum algorithms for iris dataset}
    \label{fig:iris-graph}
\end{figure}

For generalizing the proposed quantum model, we further evaluate the performance for the benchmarked Iris dataset. Similar to the results for IMDb dataset, the fully entangled quantum circuits show better performance in comparison to linearly entangled circuits. Fig. \ref{fig:iris-graph} clearly shows the efficiency of the fully entangled proposed circuit in comparison to all other quantum models and Classical SVM, where fully entangled IQP circuit shows the same results as the Classical SVM.
\section{Conclusion}
In this article, we classify the text data of IMDb movie reviews using our proposed quantum kernel based on linear and fully entangled circuits. We have also compared our results with classical machine learning algorithms and other quantum circuits available in literature. Our analysis is based on entanglement as a crucial hyper parameter for boosting the expressiveness of quantum models. Our results have further shown the significance of using different entanglement models to achieve optimum outcomes by capturing correlations between features. For five features with 10-fold cross validation, we found the fully entangled proposed circuit to be the most efficient in analysing the IMDb data- the proposed model outperforms classical support vector machine, IQP circuit and Hamiltonian circuit in addition to other classical machine learning algorithms used in this article for all evaluation metrics. For a comprehensive analysis, we further use classical support vector machine and quantum models with increased number of features to evaluate efficiencies of different algorithms. The results obtained in this article clearly demonstrate the advantages of using the proposed fully entangled circuit for a large number of features in comparison to Classical SVM and linearly entangled circuits. In fact, for higher number of features the increase in efficiency is significantly large in comparison to other algorithms. However, in comparison to linearly entangled circuits the use of Classical SVM is much more beneficial in predicting accuracy and precision. For a set of features, the proposed linearly entangled circuit also leads to precision score very close to Classical SVM. In order to validate the results obtained, we have also analysed the Iris dataset using our proposed circuits in addition to other quantum models and Classical SVM. The proposed fully entangled circuit leads to the best efficiency in terms of all evaluation metrics. For Iris data, Classical SVM and fully entangled IQP circuit led to same results. \par
Our results suggest that  the circuit depth of Hamiltonian circuit is affecting its performance as against the IQP and the proposed circuit. Therefore, a particular problem of interest would be to optimize the kernel to reduce the circuit depth with an improved efficiency of algorithms. Similar to classical machine learning algorithms where hyperparameters play a crucial role towards efficiency of a model, we expect quantum kernel models to be optimised analogously. It would be highly interesting to optimize these hyperparameters- entanglement in our case- to achieve an optimal efficiency in comparison to classical algorithms. 
\section*{Acknowledgement}
The authors acknowledge IBM for providing IBM Quantum Systems. DS further acknowledges Department of Chemistry and IDRP-QIC for providing research infrastructure, and MoE for providing the assistantship.

\bibliography{main}
\bibliographystyle{unsrt}
\end{document}